\documentclass[aps,pre,twocolumn,showpacs,superscriptaddress]{revtex4-1}  
\pdfoutput=1
\usepackage{graphicx}  % needed for figures
\usepackage{dcolumn}   % needed for some tables
\usepackage{bm,amssymb,amsmath}   % for math
\usepackage{xcolor}    % for color
\usepackage{multirow}

\begin{document}
\widetext
\title{Consensus versus persistence of disagreement\\
in opinion formation: the role of zealots\\
}
\author{Francesca Colaiori}
\affiliation{Istituto dei Sistemi Complessi (ISC-CNR), UOS Sapienza, c/o Dipartimento di Fisica, Sapienza Universit\`a di Roma, Piazzale Aldo Moro 5, 00185 Rome, Italy}
\affiliation{Dipartimento di Fisica, Sapienza Universit\`a di Roma, Piazzale Aldo Moro 5, 00185 Rome, Italy}
\author{Claudio Castellano}
\affiliation{Istituto dei Sistemi Complessi (ISC-CNR), Via dei Taurini 19, 00185 Roma, Italy}
\affiliation{Dipartimento di Fisica, Sapienza Universit\`a di Roma, Piazzale Aldo Moro 5, 00185 Rome, Italy}
\date{\today}

\begin{abstract} 
We consider a general class of three--state models where individuals
hold one of two opposite opinions, or are neutral, and exchange
opinions in generic pairwise interactions.  We show that when opinions
spread in a population where a fraction of individuals (zealots)
maintain unshakably their opinion, one of four qualitatively distinct
kinds of collective dynamics arise, depending on the specific rules
governing the social interactions.  Unsurprisingly, when their density is
high, zealots drive the whole population to consensus on their opinion. 
For low densities a rich phase diagram emerges: a finite population 
of dissenters can survive and be the only stationary state
or may need a critical mass or dissenters to be sustained; the
critical mass may vanish or not as the density is reduced; the
transition to the high density regime can be smooth or abrupt, and
shows interesting hysteretic effects. For each choice of the
interaction rules we calculate the critical density of zealots above
which diverse opinions cannot survive.
\end{abstract}
\maketitle

\section{Introduction}
Consensus often develops spontaneously in societies, leading to the adoption of widespread opinions, ideologies, traditions.  
On the other hand disagreement among individuals is also very common, even on major issues.  
Disagreement is not always due to lack of communication, and in some cases it persists even when individuals are allowed to exchange opinions and debate for long time.  
What are the mechanisms that lead to consensus, and which ingredients cause instead persistent disagreement?  
Although broadcasting media such as television and advertisement have a significant impact on individuals making their opinion, macroscopic opinion shifts in social groups may be caused by peer--to--peer interactions occurring within the social network.  
People form and reconsider their opinion while constantly interacting with others, and are often exposed to a considerable pressure to conform to the opinion of their friends and neighbors ~\cite{Castellano2009,Sen2013}. 
Yet, on certain matters, some individuals are unshakable in their opinion, and therefore totally insensitive to peer pressure.
The aim of this paper is to analyse the role of such such ``stubborn" individuals, completely reluctant to change, in determining the onset of consensus or rather the persistence of disagreement.

The effect of stubborn individuals (denoted also as committed, zealots or inflexibles) on opinion dynamics has attracted considerable interest in the past years.  
Early investigations analyzed the effect of a single~\cite{Mobilia2003} or multiple individuals~\cite{Mobilia2005,Mobilia2007} in the context of the classical voter model on lattices.  
The presence of zealots has then been studied in many other contexts, including different types of dynamics, various interaction patterns, the presence of single, multiple or competing
zealots~\cite{Galam2007,Xie2011,Marvel2012,Yildiz2013,Mobilia2013,Xie2013,Waagen2015,Svenkeson2015,Mobilia2015}.
Among the most interesting and general results is the observation~\cite{Xie2011} of a transition occurring as the fraction of zealots in the system exceeds some threshold: 
When the density of zealots is above threshold the system is driven to consensus; otherwise, alternative opinions survive in the population. 
Previous studies have generally focused on models which, in the absence of zealots, exhibit a symmetric dynamics, i.e., no opinion is a priori favored by the spontaneous evolution.  
Here we consider instead a very general class of 3--state models 
where the dynamics may either favor or disfavor the opinion held by the zealots.
In this wider context new nontrivial phenomena occur, including  the counterintuitive possibility that an opinion disfavored by the dynamics and opposed by zealots survives and is even adopted by the majority of individuals.  
In a previous paper~\cite{Colaiori2015b} we studied a similar 
general three--state opinion dynamics model 
focusing on the effect of media. 
The effect of media was schematized in the simplest possible way by assuming that people conform to the media recommendation at a constant rate and independently on their current state.
The presence of inflexible individuals considered here 
can also be interpreted, at the mean--field level, as an effective external field, but more complex than the one considered in~\cite{Colaiori2015b}:  
it induces spontaneous transitions either to the opinion supported by the media  or to the neutral state, and the transition rates depend on the current state.
The model presented here includes that of Ref. [13] as a special case. 

The paper is structured as follows. 
After the description of the general model in Section II, we write down the mean--field equation for the dynamics in Section III. 
The position and stability of their stationary solutions, depending on the model parameters,  and the consequent different shapes of the phase diagram are illustrated in Section IV, while the detailed derivation of the results is deferred to Section V.
Section VI contains a discussion of the results and concluding remarks.

\section{The model}
We analyze a model of pairwise social influence for opinion dynamics in a society consisting of two types of agents: regular agents, who update their beliefs according to the information that they receive from their social neighbors; and stubborn agents, who never update their opinion. 
Stubborn agents might for example represent political activists or customers involved in the marketing process for a 
crowdsourced campaign.
We include agents with no preferred opinion: neutral agents, who are uninformed or undecided.  
This allows significant change from models where only two opinions are allowed.  
Each agent can be in one of three states: holding opinion $A$, holding an opposing opinion $B$, or being undecided ($U$).  
A finite fraction $p$ of the population forms a sub--population $Z$ of agents committed to the opinion $A$: they never change state.  
We indicate with $n_A$, $n_B$, $n_U$ the fraction of uncommitted agents in the $A$, $B$, and $U$ state, respectively, so that $n_A+n_B+n_U+p=1$.  
The interactions rules are specified in Table~\ref{interactions}.
\begin{table}
\begin{ruledtabular}
\begin{tabular}{cccccccc} 
     &&$A,A$     & $B,B$    & $U,U$& $A,U$     & $B,U$  &$A,B$\\ 
\hline
$A-A$&&$1$        &$0$       &$0$  &$0$          &$0$     &$0$\\
$B-B$&&$0$        &$1$       &$0$  &$0$          &$0$     &$0$\\
$U-U$&&$0$        &$0$       &$1$  &$0$          &$0$     &$0$\\
$A-U$&&$\varphi_2$&$0$       &$0$  &$1-\varphi_2$&$0$     &$0$\\
$B-U$&&$0$        &$\gamma_2$&$0$  &$0$      &$1-\gamma_2$&$0$\\
\multirow{2}{*}{$A-B$}
&$A\notin Z$&
$\alpha_1$&$\alpha_2$&$\alpha_3$&$\alpha_4$   &$\alpha_5$&$\alpha_6$\\
&$A\in Z$&
$\beta_1$&$0$&$0$&$\beta_4$   &$0$  &$\beta_6$\\
\end{tabular}
\caption{Each row in the table corresponds to an interaction, and each
  column to a possible outcome.  Elements in the table indicate the
  probabilities of each possible outcome for the given interaction.
  The last two rows differ because the outcome of the $A-B$
  interaction depends on whether the agent in state $A$ is committed
  or not.  The coefficients $\alpha$ and $\beta$ are normalized:
  $\sum_{i=1}^6\alpha_i=1$, $\beta_1+\beta_4+\beta_6=1$.  Note that
  the $Z$ sub--population is constant.  }
\label{interactions}
\end{ruledtabular}
\end{table}  
Interactions involving two individuals in the same state ($A-A$, committed or not, $B-B$, $U-U$) have no effect.  
In $A-U$($B-U$) interactions undecided agents may adopt their partner's opinion with given probability: an agent holding opinion $A$($B$) has a probability $\varphi_2$ ($\gamma_2$) to convince the $U$ agent, while undecided agents do not alter the opinion of $A$($B$).  
We assume in general $\varphi_2 \ne \gamma_2$, allowing for $A$ and $B$ to have unequal persuasiveness.  
The outcome of interactions among agents holding opposite opinions ($A-B$) depends on whether the agent holding opinion $A$ is committed or not.  
When $A$ is uncommitted ($A\notin Z$), the interaction may have any outcome: each of the two agents may either keep her opinion, change it to match the opinion of the partner, or get confused and turn to the undecided state, in any combination.  We indicate with $\alpha_i$, $\{i=1,6\}$ ($\sum_{i=1}^6\alpha_i=1$) the probabilities of the six possible outcomes  (see Table~\ref{interactions}).  
Interactions between an agent committed to opinion $A$ ($A\in Z$) and an agent holding opinion $B$ have instead only three possible outcomes, corresponding to the three possible states of $B$ after the interaction, since the state of $A$ is unaltered.  
We indicate with $\{\beta_i$, $i=1,4,6\}$ ($\beta_1+\beta_4+\beta_6=1$) the probabilities of these outcomes (see Table~\ref{interactions}).

\section{The dynamics on a complete graph}
We now write the equations for the dynamics of this general model on a complete graph of infinite size (i.e. in mean-field),  describing the time evolution of the density of uncommitted agents in each state:
\begin{equation}
\!\!\left\{
\begin{array}{lll} 
\!\! \dot{n}_A&
\!\!=\!\!\!\!&
2\beta_1 p n_B+2\varphi_1 n_A n_B +2 \varphi_2 (n_A+p)n_U\!\!\!\!\\
\!\!\dot{n}_B &
\!\!=\!\!\!\!&
-2(\beta_1+\beta_4)p n_B+2\gamma_1 n_A n_B  +2 \gamma_2 n_B n_U  
\end{array}
\!\!\!\right.
\label{dynamics_z}
\end{equation}
where $\varphi_1=\alpha_1-\alpha_2-\alpha_3-\alpha_5$ and $\gamma_1=-\alpha_1+\alpha_2-\alpha_3-\alpha_4$ represent the net gain in $A$ ($B$) states in an $A-B$ interaction.  
The equation for $\dot{n}_U$ can be derived from the normalization condition.  
Defining $r=p/(1-p)$, and $\tilde{n}_i=n_i/(1-p)$ ($i=A,B,U$) as the fraction of uncommitted agents respectively in the $A$, $B$ and $U$ state, normalized with respect to the total uncommitted population ($\tilde{n}_A+\tilde{n}_B+\tilde{n}_U=1$), Eqs.~(\ref{dynamics_z}) can be rewritten as:
\begin{equation}
\!\!\!\!\left\{\begin{array}{lll} 
\!\!\dot{\tilde{n}}_A(1+r)&
\!\!=\!\!\!\!&
2r\beta_1\tilde{n}_B+2r\varphi_2\tilde{n}_U+2\varphi_1 
\tilde{n}_A\tilde{n}_B+2\varphi_2\tilde{n}_A\tilde{n}_U\!\!\!\\
\!\!\dot{\tilde{n}}_B (1+r)&
\!\!=\!\!\!\!&
-2r (\beta_1+\beta_4)\tilde{n}_B+2\gamma_1 \tilde{n}_A 
\tilde{n}_B+2\gamma_2\tilde{n}_B \tilde{n}_U
\end{array}\!\!\!\right.
\label{dynamics1}
\end{equation}
From Eqs.~(\ref{dynamics1}) it is clear that, in terms of the the densities of uncommitted sub--populations and with a proper rescaling of the time variable, the system maps exactly onto a system of uncommitted agents in an external field acting as an exogenous one--to--many brodcasting source.  
Interactions and the effect of the external field are specified as follows: The external field induces individuals holding opinion $B$ to switch to opinion $A$ at rate $r \beta_1$, and to become undecided ($U$) at rate $r \beta_4$; undecided individuals acquire opinion $A$ at rate $r \varphi_2$.  
The interactions are those described in Table~\ref{interactions} for uncommitted agents.  
A similar case was studied in~\cite{Colaiori2015b} where the focus was on the interplay between external media and interpersonal influence events occurring within the social network. 
In that case the effect of media was schematized in the simplest possible way by assuming that people conform to the media recommendation at a constant rate and independently on their current state.  
The model presented here can be seen as a generalization of Ref.~\cite{Colaiori2015b}.  
The presence of 
%zealot individuals 
zealots introduces in the dynamical equations additional terms, that can be interpreted as external fields inducing the spontaneous transitions $B\to A$ and $U \to A$ (but at {\em different} rates), as well as the transition $B \to U$.  
The model studied in Ref.~\cite{Colaiori2015b} is recovered as a special case of the model studied here, with $\beta_1=\varphi_2$, $\beta_4=0$, and $r \rightarrow r/2\tilde{\varphi}_2$, so that $r_{B\rightarrow A}=r$, $r_{B\rightarrow U}=0$, $r_{U\rightarrow A}=2\varphi_2 p =r$.  
For $r=0$ the two models coincide.

\section{Collective behavior at stationarity}
\begin{figure*}
\centering
\includegraphics[width=\textwidth]{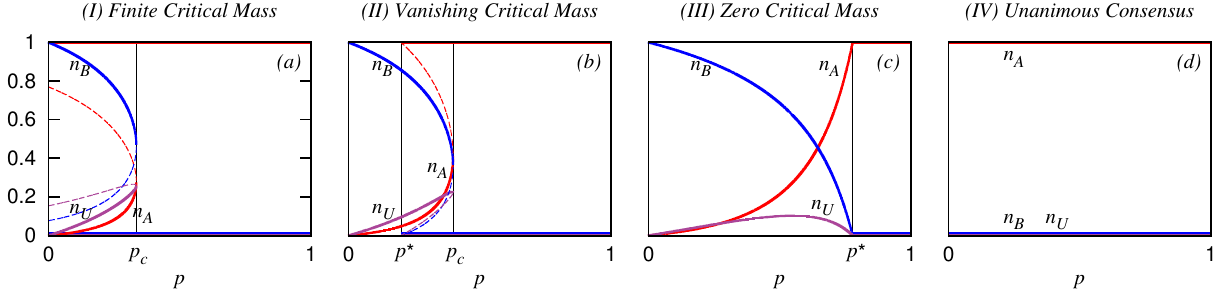}
\caption{(color online) Theoretical results for the densities of
  agents for realizations of each of the four classes of models  (a)
  {\em Class I} model (FCM) ($\varphi_1=-0.8$, $\varphi_2=0.4$,
  $\gamma_1=-0.1$, $\gamma_2=0.5$, $\beta_1=0.1$, and $\beta_4=0.1$), (b){\em Class II} model (VCM)
  ($\varphi_1=-0.8$, $\varphi_2=0.4$, $\gamma_1=0.05$, 
  $\gamma_2=0.5$, $\beta_1=0.1$, and $\beta_4=0.1$), (c){\em Class III} model (ZCM) ($\varphi_1=-0.8$,
  $\varphi_2=0.4$, $\gamma_1=0.7$, $\gamma_2=0.5$, $\beta_1=0.1$, and $\beta_4=0.1$), (d){\em Class
    IV} model (TC) ($\varphi_1=0.5$, $\varphi_2=0.4$, $\gamma_1=-0.5$,
   $\gamma_2=0.5$, $\beta_1=0.1$, and $\beta_4=0.1$).  Solid (dashed) lines represent stable
  (unstable) lines. 
  }
\label{Densities}
\end{figure*}
\begin{figure}
\centering
\includegraphics[width=8.5cm]{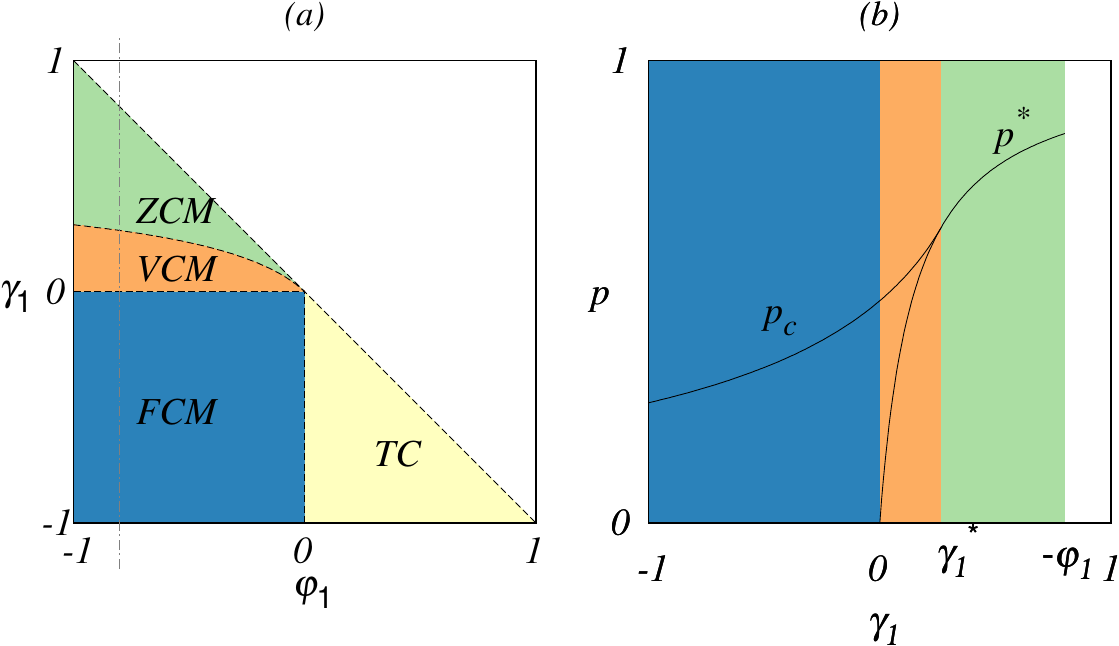}
\includegraphics[width=8.5cm]{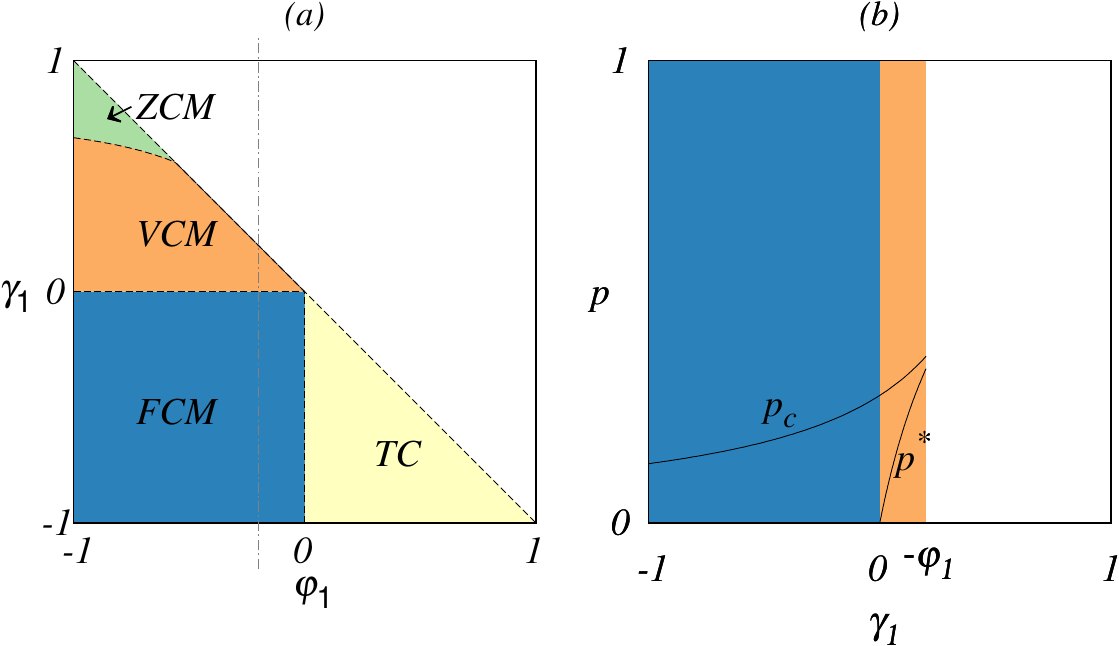}
\caption{(color online)(a,c) Phase diagram in the ($\varphi_1$,
  $\gamma_1$) plane.  The curve $\gamma_1^{*}$ separating regions
  $VCM$ and $ZCM$ depends on $\varphi_2$, $\gamma_2$, $\beta_1$ and $\beta_4$ In the upper panel (a) $\varphi_2=0.4$, $\gamma_2=0.7$, $\beta_1=.2$, and $\beta_4=.2$, so that $\varphi_2>\gamma_2 \beta_4/(\beta_1+\beta_4)$. In the lower panel (c) $\varphi_2=0.05$, $\gamma_2=0.8$, $\beta_1=.2$, and $\beta_4=.2$, so that $\varphi_2<\gamma_2 \beta_4/(\beta_1+\beta_4)$.  (b,d) Plot of $p_c$ (saddle--node bifurcation curve) and $p^*$ (transcritical bifurcation line) as functions of $\gamma_1$, and for
  $\varphi_1=-0.8$ (upper panel, (b)), $\varphi_1=-0.2$ (lower panel, (d)).}
\label{PD}
\end{figure}

Choosing $x=\tilde{n}_B$ and $y=\tilde{n}_A$ as independent variables, the stationarity condition in Eq.~(\ref{dynamics1}) reads:
\begin{equation}
\left\{\begin{array}{lll}
y^2-y(1-r+\varphi x)+r \alpha x-r = 0 ,& \,\,\,\,&{\cal C}_1 \\
x(\gamma y- x +1- r\beta)=0  &  &{\cal C}_2 \,, 
\end{array}\right.
\label{curves2}
\end{equation}
where $\varphi=(\varphi_1-\varphi_2)/\varphi_2$,
$\gamma=(\gamma_1-\gamma_2)/\gamma_2$,
$\alpha=\varphi_2-\beta_1)/\varphi_2$, and
$\beta=(\beta_1+\beta_4)/\gamma_2$, and we assume $\varphi_2\neq0$,
$\gamma_2\neq 0$.
The stationary solutions of the dynamic equations are therefore given by the intersections of the two conic sections ${\cal C}_1$ and ${\cal   C}_2$ that fall within the physical region, which is the triangle delimited by the lines $x=0$, $y=0$ and $y=1-x$.  
The first conic section, ${\cal C}_1$, is a non--degenerate hyperbola, while the second one, ${\cal C}_2$, factorizes into the product of the two straight lines:
\begin{equation}
%\left\{
\begin{array}{lll}
x=0 &\,\,\,\,& {\cal R}_1\\
y=\gamma^{-1} x+\gamma^{-1}\left(r\beta-1\right)&& {\cal R}_2 \,.
\end{array}
%\right.
\label{lines}
\end{equation}

Only one of the two intersections of ${\cal R}_1$ with ${\cal C}_1$, the one at $P_1=(0,1)$, falls within the physical region; it is a stationary solution for any set of parameters values and for any $r$. $P_1$ corresponds to the trivial absorbing state of unanimous consensus on opinion $A$.  
The collective behavior of the system for any given set of parameters is fully determined by the possibility that other non--trivial fixed points (any possible intersections of ${\cal C}_1$ and ${\cal R}_2$) appear within the physical region as the control parameter $r$ is varied.  
The existence, position and stability of non--trivial fixed points depend on the specific values of the six parameters ($\gamma_1$, $\gamma_2$, $\varphi_1$ $\varphi_2$, $\beta_1$, $\beta_4$) defining the model.

In what follows we classify each model, specified by a set of parameters, according to its collective behavior as the control parameter $r$ is varied ($r$ is a monotonic function of the fraction $p$ of committed agents). 
 The behavior at large $r$ is independent of the parameters values; trivially, in that regime, the full consensus on opinion $A$ is the only stable stationary state.
However, in the small $r$ regime, we observe four qualitatively distinct types of response to the variation of the control parameter.
We therefore classify all possible specific models, each one specified by a set of parameters, in four classes.  
The phase diagram for each class is shown in Fig.~\ref{Densities}, while the ``meta--phase diagram" giving the classification of the models according to their set of parameters is summarized in Fig.~\ref{PD}.  
We now describe the phenomenology for each one of the four classes, and postpone the detailed analysis of the solutions of Eqs.~(\ref{curves2}) to the next section.

\subsection*{Class I: Finite Critical Mass (FCM) Models} 
When $\varphi_1<0$, $\gamma_1 \leq 0$, i.e. in models where $A-B$ interactions produce on average an increase in undecided individuals, with no net gain in $A$ nor in $B$ states, the system undergoes a first order transition at a finite value $r=r_c$ of the external bias [see Fig.~\ref{Densities}(a)].  
For large enough fraction of committed (large $r$), $P_1$ is the only fixed point and the system flows into the absorbing state of total consensus on opinion $A$, for any initial condition. 
 At $r=r_c$ the system undergoes a saddle--node bifurcation~\cite{Strogatz2001}: one double solution appears.  
 As $r$ is decreased below $r_c$ the double solution splits into one stable (the one with larger $n_B$), and one unstable solutions.  
 In the nontrivial stable fixed point the two opinions $A$ and $B$ coexist in the population, together with a fraction of undecided [see   Fig.~\ref{Densities}(a)].  We call this state, where disagreement persists, ``pluralism". 
  The initial conditions determine whether consensus is asymptotically reached ($n_A=1$), or disagreement persists ($n_B>0$).  
  The value of $n_B$ at the unstable fixed point stays finite in the limit $r\rightarrow 0$, implying that, no matter how small is the fraction of individuals committed to opinion $A$, a finite ``critical mass"~\cite{Dodds2004} of dissenters (supporting opinion $B$ in the initial state) is always necessary to reach the pluralistic state.

\subsection*{Class II: Vanishing Critical Mass (VCM) Models}
This class is identified by the range of parameters $\varphi_1<0$, $0< \gamma_1 < \min(\gamma_1^*,-\phi_1)$  [see below for the definition of $\gamma_1^*$],  and corresponds to models where $A-B$ interactions cause on average a small increase of $B$ states at the expense of $A$ states. 
 As in {\em Class I} models, lowering $r$ below $r_c$ the system undergoes a first order transition separating a regime ($r > r_c$), where consensus on opinion $A$ is the only stable state from a regime ($r \leq r_c$) where a stable and an unstable additional fixed points appear through a saddle--node bifurcation (see next section).  
 However, in this case, further reducing $r$, the unstable fixed point collides with the point $P_1$ (consensus on $A$) at a finite value $r^*$ ($0<r^*<r_c$), and then exits the physical region.  
 Correspondingly, the value of $n_B$ at the unstable fixed point vanishes when $r \to r^*$.  This is a transcritical bifurcation~\cite{Strogatz2001}: when the two fixed points cross each other, they exchange stability; $P_1$ becomes unstable, so that below $r^*$ the system, unless started with $n_B \equiv 0$, always flows to the pluralistic state [Fig.~\ref{Densities}(b)].  
 Therefore in this case, the initial presence of even a few dissenters suffices for
opinion $B$ to survive.  
The curve $\gamma_1=\gamma_1^*$ separating {\it Class II} and {\it III} depends on all other model parameters.
In a certain range of parameters $\gamma_1^*$ becomes larger than $-\varphi_1$ so that all physical values of $\gamma_1$ are always below  $\gamma_1^*$, and {\it Class III} disappears [see Fig.~\ref{PD}(c)].
The transcritical bifurcation line is $r=r^*=\gamma_1/(\beta_1+\beta_4)$, and always lies below the saddle--node bifurcation line.

\subsection*{Class III: Zero Critical Mass (ZCM) Models}
When $\varphi_1<0$, $\gamma_1 \geq \gamma_1^*$, corresponding to models where $A-B$ interactions give an increase of undecided and a large increase in $B$ at the expense of $A$, the system undergoes, at $r=r^*$, a continuous transition (transcritical bifurcation) between total consensus on opinion $A$ ($r>r^*$) and pluralism ($r<r^*$), see Fig.~\ref{Densities}(c).  In this class of behavior initial conditions do not play any role.

\subsection*{Class IV: Total Consensus (TC) Models}
This class corresponds to the region $\varphi_1\geq 0$, i.e. to models where $A-B$ interactions result in a net increase of individuals holding opinion $A$.  
The behavior is trivial: irrespectively of the value of all other parameters, for any initial condition, and no matter how small the density $p$ of committed individuals is, the system always converges to the consensus state ($P_1$) [see Fig.~\ref{Densities}(d)].

\section{Derivation of the phase diagrams}
We now show how the rich phenomenology described in the previous section and summarized in Fig.~\ref{Densities} is derived from Eqs.~(\ref{dynamics_z}).  
We need to find the solutions of Eqs.~(\ref{curves2}) and look at how their behavior, as the control parameter $r$ (or equivalently $p$) varies, depends on the parameters
defining the model.  
Depending on the values of the six parameters ($\gamma_1$, $\gamma_2$, $\varphi_1$, $\varphi_2$, $\beta_1$, $\beta_4$) controlling the interactions, different collective
behaviors emerge.

\begin{figure}
\includegraphics[width=8.cm ]{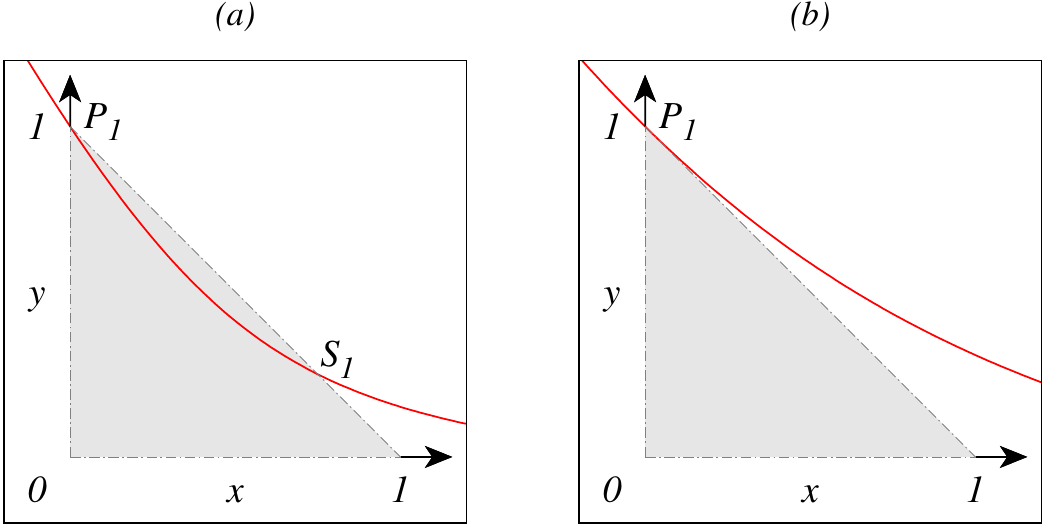}
\caption{(color online) Sketch of the qualitative change of the curve 
${\cal C}_1$ (solid line) for $\varphi_1\leq0$ and $0<r\leq \tilde{r}$
[panel (a)] and for all other cases [panel (b)]. The grey shaded domain is
the physical region.}
\label{PhysReg}
\end{figure}

In order to find the intersections of ${\cal C}_1$ with ${\cal R}_1$ and ${\cal R}_2$ that correspond to physically meaningful fixed points, it is useful to first identify the position of ${\cal C}_1$ with respect to the physical region. This can be done by locating the intersections of ${\cal C}_1$ with the boundary of the physical region, delimited by the three lines $x=0$, $y=0$, $y=1-x$.  
{\it (i)} The intersections of ${\cal C}_1$ with $x=0$ are in $P_1=(0,1)$ and
$P_2=(0,-r)$; only the first one belongs to the physical region.  
{\it (ii)} The only intersection of ${\cal C}_1$ with $y=0$ is in $(\alpha^{-1}, 0)$, which, since $\alpha\leq 1$, is always outside the physical region.  
{\it (iii)} The intersections of ${\cal C}_1$ with $y=1-x$ are in $P_1=(0,1)$ and $S_1=(1-r/\tilde{r},r/\tilde{r})$, with $\tilde{r}=(\varphi+1)/(\alpha-1)=-\varphi_1/\beta_1$; $S_1$ belongs to the physical region only for $\varphi_1\leq0$ and $r\leq \tilde{r}$.
Note that whenever $S_1$ belongs to the physical region (for $\varphi_1\leq0$ and $r\leq \tilde{r}$) the upward concavity of ${\cal C}_1$ guarantees that ${\cal C}_1$ actually enters the physical region.  
The shape of ${\cal C}_1$ is qualitatively represented in Fig.~\ref{PhysReg}, where (a) corresponds to $\varphi_1\leq0$ and $0<r\leq \tilde{r}$, and (b) to all other cases where ${\cal C}_1$ intersects the physical region in just one point ($P_1$).

$P_1=(0,1)$ always belongs both to ${\cal C}_1$ and to ${\cal R}_1 $
($P_1$ is in fact the only intersection of ${\cal R}_1$ with the upper
branch of ${\cal C}_1$), therefore $P_1$ is a fixed point for all
possible choices of the parameters and for any $r$. It corresponds to
the trivial absorbing state of unanimous consensus on opinion $A$.
For $\varphi_1 \geq 0$ (corresponding to {\em Class IV} models) ${\cal
  C}_1$ intersects the physical region only in ${P_1}$ [see
  Fig.~\ref{PhysReg}(b)], thus in this case there cannot be other
fixed points besides $P_1$. Thus we restrict the following analysis to
the non--trivial case $\varphi_1 <0$: in this case, while $P_1$ still
is the only fixed point for sufficiently large fraction of committed
agents [$r\geq\tilde{r}$, see Fig.~\ref{PhysReg}(b)], additional fixed
points may arise for $r< \tilde{r}$ from the intersections of ${\cal R}_2$ 
and ${\cal C}_1$ within the physical region.  

\subsubsection*{$r=0$}

\begin{figure}
\includegraphics[width=8.cm ]{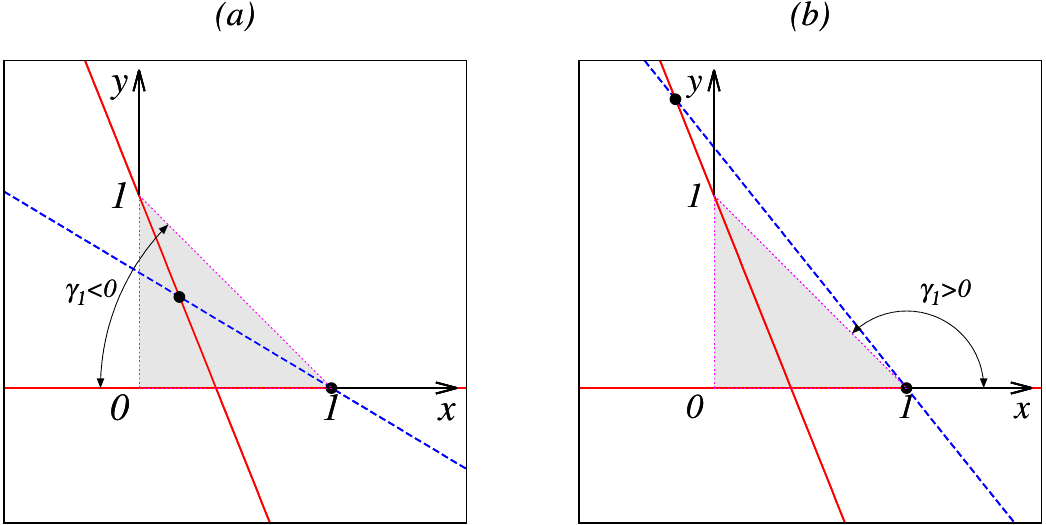}
\caption{(color online) Different qualitative behaviors of the stationary
solutions for $r=0$. The red solid line is the curve ${\cal C}_1$. The
blue dashed line is ${\cal R}_2$. (a) For $\gamma_1>0$ the intersection $Q_2$
of the two curves falls outside the physical region (grey-shaded). (b)
For $\gamma_1<0$ $Q_2$ is instead within the physical region. The black
curved lines
indicate the range of variability of the slope of ${\cal R}_2$ as $\gamma_1$
is changed.}
\label{Fig_1}
\end{figure}

To understand how the fixed points move, as the control parameter $r$
is varied, it is useful to first understand what happens in the case
$r=0$, that is the limit of vanishing density of committed agents.
For $r=0$ the line ${\cal R}_2$ always goes
through $Q_1=(1,0)$, that also belongs to the asymptote $y=0$ of ${\cal C}_1$,
and is therefore a fixed point.  It is clear that, since the slope of ${\cal
  R}_2$ is $\gamma^{-1}$, for $\gamma>-1$ ($\gamma_1>0$) $Q_1$ is the
only point of ${\cal R}_2$ that falls within the physical region (see 
Fig.~\ref{Fig_1}).
Let us rewrite ${\cal C}_1$ as
\begin{equation}
%(y-r\alpha/\varphi)(y-\varphi x -1 +r(1+\alpha/\varphi))=r(1+\alpha/\varphi)(1-r\alpha/\varphi)
(\varphi y-r\alpha)(\varphi y-\varphi^2 x -\varphi +r(\varphi+\alpha))=r(\varphi+\alpha)(\varphi-r\alpha)\label{C1_asy}
\end{equation}
that allows to identify the asymptotes.  In the limit $r\rightarrow 0$
${\cal C}_1$ factorizes in the product of its asymptotes, that, in the
same limit, are $y=0$, and $y=\varphi x +1$.  Note that $\varphi_1<0$
implies $\varphi<-1$, so that both asymptotes cross the physical
region at $r=0$.  When $r=0$ the intersections of ${\cal C}_1$ with
${\cal R}_2$ are $Q_1=(1,0)$ (consensus on opinion $B$, always within
the physical region), and
$Q_2=((1+\gamma)/\Gamma,(1+\varphi)/\Gamma)$, where
$\Gamma=1-\gamma\varphi$.  $Q_2$ falls within the physical region as
long as $\gamma_1 \leq 0$ (that implies that $1+\gamma$ and
$1+\varphi$ have the same sign, and therefore $Q_2$ belongs to the
first quadrant), see Fig.~\ref{Fig_1}(a). 
$Q_2$ coincides with $P_1$ for $\gamma_1=0$. For
$\gamma_1>0$ $Q_2$ is always outside the physical region (to the left
of $P_1$ for $\Gamma<0$ [see Fig.~\ref{Fig_1}(b)] or to the right of 
$Q_1$ for $\Gamma>0$).
For $\gamma>-1$ ($\gamma_1>0$) 
there are overall three fixed points in the physical region: 
$P_1$ (unanimous consensus on
opinion $A$) at the intersection of ${\cal C}_1$ with ${\cal R}_1$;
$Q_1$ (unanimous consensus on opinion $B$) at the intersection
of ${\cal C}_1$ with ${\cal R}_2$; $(0,0)$, given by the other
intersection of ${\cal C}_1$ with ${\cal R}_1$. 
For $\gamma<-1$ ($\gamma_1<0$) instead, ${\cal R}_2$ always crosses the physical
region, and intersects the other asymptote in some $Q_2$ inside the
physical region, giving another fixed point, besides the two
corresponding to unanimous consensus, and the one in $(0,0)$.  Note
that for $r=0$ the fixed point in $(0,0)$ always is unstable, and
disappears as soon as $r>0$.

\subsubsection*{$r>0$}
\begin{figure}
\includegraphics[width=8.cm ]{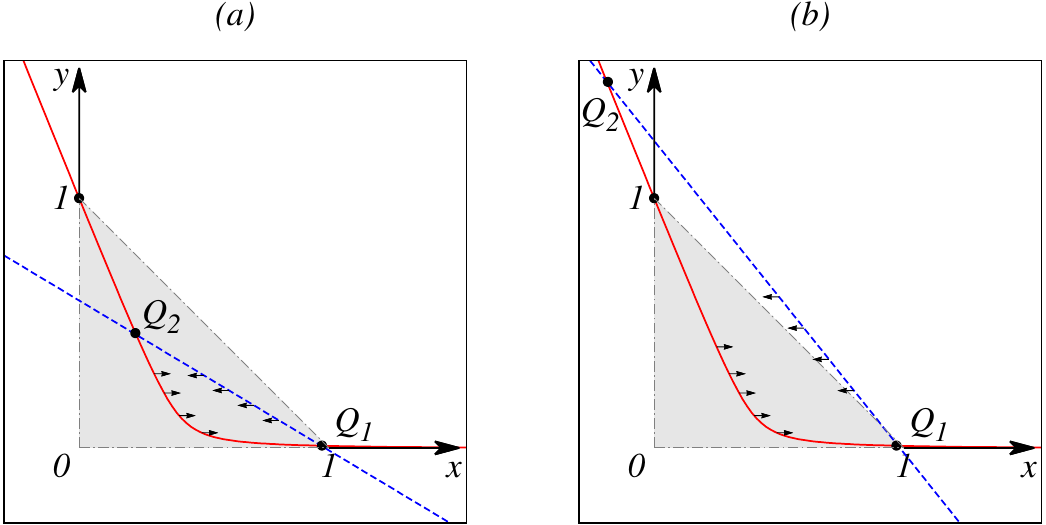}
\caption{(color online) Sketch of the motion of the curves 
${\cal C}_1$ (red solid curve) and ${\cal R}_2$ (blue dashed line) when $r$ 
is increased. (a) The intersection point $Q_1$ moves leftward, while $Q_2$
moves rightward. For $r=r_c$ they collide and annihilate. (b) The intersection point $Q_1$ moves leftward, while $Q_2$ moves rightward. They may collide and annihilate either inside or outside the physical region. 
}
\label{Fig_2}
\end{figure}

As $r$ increases, ${\cal R}_2$ always moves leftward (the slope does
not depend on $r$ and the intersection with the $x$ axis
is $(1-r\beta,0)$ always moves leftward as $r$ increases).  
In the meantime, the upper branch of ${\cal C}_1$ moves 
rightward~\cite{nota0}. 
Therefore, the intersection $Q_1$ always moves leftward. 
We must now distinguish two cases: \\
1) ${\cal C}_1$ and ${\cal R}_2$ having one
intersection on each branch: this occurs for
$\Gamma=1-\gamma\varphi>0$. In this case the situation is simple:
$Q_1$ moves leftward as $r$ increases and for some value $r=r^*$
reaches $P_1$ and then leaves the physical region. The other
intersection $Q_2$ is always on the unphysical branch of ${\cal C}_1$
(the intersection between ${\cal R}_2$ and the line $y=1-x$, located
in $(1-r/\tilde{\tilde{r}},r/\tilde{\tilde{r}})$, with
$\tilde{\tilde{r}}=\gamma_1/(\beta_1+\beta_4)$ is always to the left
of $S_1$, since $\tilde{\tilde{r}}\leq\tilde{r}$ follows from the
physical constraint $\gamma_1+\varphi_1<1$, therefore $Q_1$ can only
exit the physical region crossing $P_1$).  \\
2) ${\cal C}_1$ and
${\cal R}_2$ having two intersections on the same branch: this occurs
for $\Gamma=1-\gamma\varphi<0$ (and small enough $r$)
(see Fig.~\ref{Fig_2}). In this case,
${\cal R}_2$ always has two intersections with ${\cal C}_1$ for $r=0$:
$Q_1$ at $(1,0)$ that moves to the left as $r$ increases, and $Q_2$,
that is within the physical region for $\gamma_1<0$, and outside the
physical region (to the left of $P_1$) when $\gamma_1>0$. In both
cases, as $r$ increases $Q_1$ moves leftward, and $Q_2$ moves
rightward, and they collide at some point for some $r=r_c$.
When $\gamma_1<0$ ($Q_2$ within the physical region for $r=0$) $Q_1$
and $Q_2$ collide for some finite $r=r_c$. %(then there is no intercept
for $r>r_c$.  In the other case, $\gamma_1>0$, again, as $r$
increases $Q_1$ and $Q_2$ get closer and closer ($Q_1$ moving
leftward, and $Q_2$ moving rightward), but two different events may
happen depending on the value of $\gamma_1$: $Q_1$ and $Q_2$ either 
collide within the physical region ($Q_2$ crosses $P_1$ for some $r=r^*$
entering the physical region, and then $Q_2$ collides with $Q_1$ for
some $r_c>r^*$), or they collide outside the physical region ($Q_1$
crosses $P_1$ for some $r=r^*$, exiting the physical region and then
$Q_1$ collides with $Q_2$ at some unphysical point to the left of
$P_1$ for some $r_c>r^*$).  A special value
$\gamma_1=\gamma_1^*$ separates the two behaviors: $\gamma_1^*$ is the
value such that $Q_1$ and $Q_2$ collide exactly in $P_1$, on the
boundary of the physical region.

\subsubsection*{Derivation of $r^*$ and $\gamma_1^*$}

To obtain the critical value $r^*$, we observe that $r^*$ is such that
at least one intersection between ${\cal C}_1$ and ${\cal R}_2$
coincides with $P_1$, which requires that ${\cal R}_2$ goes through
$P_1$, and therefore
\begin{equation}
r^*=\gamma_1/(\beta_1+\beta_4)\,.
\label{r*}
\end{equation}
A critical value $r_c$ such that ${\cal C}_1$ and ${\cal R}_2$ have a
double intersection on the upper branch always exist under the
necessary condition $\Gamma<0$, however it is physically meaningful
only as long as such double intersection falls within the physical
region. We have seen that $Q_1$ and $Q_2$ certainly collide within the
physical region for $\gamma_1<0$, and for small enough positive
$\gamma_1$ ($0<\gamma_1<\gamma_1^*$). $\gamma_1^*$ is defined as the
value of $\gamma_1$ such that $Q_1$ and $Q_2$ collide in $P_1$, on the
boundary of the physical region.
Imposing this condition we find that the value of $r_c$ at $\gamma_1^*$ is $r_c=r^*=\gamma_1^*/(\beta_1+\beta_4)$, and that $\gamma_1^*$
is the positive solution of 
\begin{equation}
%\alpha\gamma^{*2}+\gamma^*(1+\alpha-\varphi\beta)+1+\beta=0\,.
\alpha(\gamma_1^{*}/\gamma_2)^2
+\gamma_1^{*}/\gamma_2 (\varphi\beta+\alpha-1)
+\beta(1+\varphi)=0\,.
\label{gammastar}
\end{equation}

One further piece is missing: while for any given $\varphi_1$ a
positive value of $\gamma_1^*$ always exists, it could result to be
unphysical in certain ranges of the other parameters $\varphi_2$,
$\gamma_2$, $\beta_1$, and $\beta_4$. Indeed, besides being positive,
physically meaningful values of $\gamma_1$ have to satisfy the
condition $\gamma_1+\varphi_1\leq1$. Imposing that
$\gamma_1^*<-\varphi_1$ and after some lengthy but trivial calculation
we find the condition $\varphi_2\geq
\gamma_2\beta_4/(\beta_1+\beta_4)$. In the parameter region $\varphi_2
<\gamma_2\beta_4/(\beta_1+\beta_4)$, $\gamma_1^*$ becomes larger than
the upper bound $-\varphi_1$, therefore physically accessible values
of $\gamma_1$ are always smaller than $\gamma_1^*$, and the
corresponding models are in {\em Class III} for any physical positive
$\gamma_1$ ($0<\gamma_1<-\varphi_1$).

\subsubsection*{Wrap--up: Building the  phase diagram from the fixed points}
We now have all the elements to summarize the different classes of collective behavior, that depend crucially on the parameters $\varphi_1$ and $\gamma_1$.  
For large $r$ the trivial, stable absorbing state of unanimous consensus on opinion $A$  ($P_1$) is the only fixed point for all possible sets of parameter values.  
When $\varphi_1>0$ the state of unanimous consensus on opinion $A$ is the only fixed point for any $r$ ({\em TC} models).
When $\varphi_1\leq0$, we can distinguish three qualitative different behaviours according to the value of $\gamma_1$: {\it (i)} For $\varphi_1<0$, $\gamma_1<0$ ({\em FCM} models), as $r$ becomes smaller than a critical value $r_c$ two additional fixed points are formed by means of a saddle--node bifurcation (discontinuous transition).  
The two fixed points remain physical for any value of $r$ down to $r=0$.
{\it (ii)} For $\varphi_1<0$, $0\leq \gamma_1 \leq \min(\gamma_1^*,-\varphi_1)$ ({\em VCM} models) there is a discontinuous transition at $r=r_c$.  
When $r$ is further reduced one of the fixed points collides with $P_1$ for $r=r^*$ and then exits the physical region.  
At $r=r^*$ the fixed point $P_1$ becomes unstable and the non--trivial fixed point $Q_2$ remains the only stable stationary state, unavoidably reached by the dynamics. 
{\it (iii)} For $\varphi_1<0$, $\min(\gamma_1^*,-\varphi_1) \leq \gamma_1 \leq -\varphi_1$ ({\em ZCM} models), at $r=r^*$ the point $Q_1$ enters the physical region, thus
becoming stable, while $P_1$ turns unstable.  
A continuous transition occurs between unanimous consensus on $A$ (for $r>r^*$) and a pluralistic state with $\tilde{n}_B>0$ (for $r<r^*$).

For $\varphi_2 < \gamma_2\beta_4/(\beta_1+\beta_4)$ $\gamma_1^*$ is always below $-\varphi_1$ corresponding to the phase diagram in Fig~\ref{PD}(a).  When $\varphi_2 \geq
\gamma_2\beta_4/(\beta_1+\beta_4)$ instead $\gamma_1^*$ crosses $-\varphi_1$ at $\varphi_1=\varphi_1^*= -(\gamma_2\beta_4-\varphi_2(\beta_1+\beta_4)) /(\varphi_2+\beta_4)$. 
In this case, for $\varphi_1\geq \varphi_1$ there is no continuous transition, and the phase diagram is as in Fig~\ref{PD}(c).

\section{Discussion and Conclusions}
In this paper we have introduced and solved analytically (in mean--field) a very general model for opinions dynamics in the presence of committed agents.
 We have shown that when opinions spread in a population  where a given fraction of individuals have unshakable opinions, four qualitatively distinct kinds of collective dynamics emerge, depending on the specific rules governing the social interactions. 
 We have categorized this very extensive set of opinion dynamics models accordingly within mean--field.
To keep the interaction rules very general the models considered depend on a large number of parameters. 
A high density of zealots trivially drives the system towards total consensus, irrespectively of all the parameters ruling the interactions.  
The qualitative behavior of the stationary states at smaller densities depends essentially only on the pair ($\gamma_1$, $\varphi_1$), which encode information about interactions among individuals with opposite opinions. The other parameters have a role in locating the tricritical line, the saddle--point bifurcation line and the transcritical bifurcation line.  
In some parameters regions disagreement persists, while in others consensus is achieved. The final state could depend or not on initial conditions, and the behavior at low densities can change smoothly or abruptly to that at high $p$. In one parameters region 
consensus is always achieved also at low densities.    
Our results allow to precisely predict when these different behaviors occur.

Note that, although the parameters regulating the interactions with undecided agents have a marginal role in determining the collective behavior, the presence itself of the $U$ state is crucial for a nontrivial behavior~\cite{nota3,Colaiori2015}. 

Above some critical density of zealots, the system always reaches asymptotically a state of unanimous consensus. 
The stability of the consensus obtained upon fluctuations of the zealots density depends on the specific model. For models the FCM class consensus is very stable: once it is reached, it is maintained  when the density of zealots is diminished, even down to zero. In models in the ZCM class dissenters nucleate as soon as the density of zealots fluctuates below the threshold. 
Models in the VCM class shows an interesting hysteretic behavior: Once consensus is reached with a sufficiently high zelots density, it is maintained when the density goes below threshold, however, large fluctuations below $p_c$ make the system unstable, and at $p^*$  any infinitesimal perturbation causes an abrupt transition to a state with a finite density of $B$ states.

We finally note a counterintuitive behavior that emerges in certain parameters ranges, where $B$ states  survive, and may even become the majority, although both the effect of the zealots and the interaction rules are biased against them~\cite{nota4}. 
The role of the neutral state is crucial in producing such behavior: Although the peer interaction rules favor the $A$ state, the rate at which they occur allows the $B$ state to be favored on average~\cite{Colaiori2015}.

For every single choice of the parameters our analysis identifies the value of the critical density of zealots ($\overline{p}=\max(p_c, p^*)$ , with $p_c=r_c/(1+r_c)$ and $p^*=r^*/(1+r^*)$) above which diverse opinions cannot survive.   
This critical fraction of committed individuals needed to achieve consensus has an upper bound in $\tilde{p}=\tilde{r}/(1+\tilde{r})=1/(1-\beta_1/\varphi_1)$. 
In order for this bound to be non--trivial it has to be $\varphi_1<0$ (indeed for $\varphi_1>0$ the system always reaches the consensus state). The bound $\tilde{p}$ only depends on the ratio $\beta_1/\varphi_1$ that measures the relative net gain (or loss) of individuals in an $A$ state after an $A-B$ interaction when $A$ is committed with respect to the case when $A$ is uncommitted. 
When the average net gain in $A-B$ interactions with committed $A$ is larger then the average net loss in $A$ due to $A-B$ interactions with uncommitted $A$, i.e. for $\beta_1>-\varphi_1$ the bound ensures  $\tilde{p}<1/2$, meaning that a committed minority suffices to drive the system to consensus.  

In our analysis, we have considered generic rules for the peer interactions, but we assumed symmetric roles for the two interacting partners. However, we point out that our analysis holds more generally, including cases where the interaction partners have distinct roles (e.g., speaker--listener), as often considered in the literature. This can easily be shown along the same lines followed in Ref.~\cite{Colaiori2015b}, where we have shown that  model allowing for asymmetric roles is always equivalent at the mean--field level to its symmetrized version, the outcome of an interaction in the symmetrized model being defined as the average result of two asymmetric interactions with exchanged roles.

Finally, we remind that our results describe stationary properties of the general model under investigation. The time scales needed to reach the steady state may be strongly dependent on the different parameters in Eqs.~(\ref{dynamics1}). Moreover, it should be noted that, since only $A$ zealots are present, consensus on the $A$ opinion is the only possible stationary state in a system of finite size. All other mean--field stationary solutions may appear only as quasi--stationary steady--states.

The present investigation allows to recover, in a unified framework, various types of behavior previously uncovered in several other modeling attempts, thus allowing to understand the physical ingredient behind the single observations.
In particular some of the dynamics considered in Ref.~\cite{Marvel2012}, are special cases of our general model. 
For example, the ``basic model'' of Ref.~\cite{Marvel2012} correponds to  $\gamma_1=\varphi_1=-1/2$,  $\gamma_3=\varphi_2=1/2$, $\beta_1=0$, $\beta_4=1/2$. 
From our analytical treatment, it is immediate to conclude that such a model exhibits a discontinuous transition as $p$ is decreased from unanimous consensus to coexistence of individuals supporting both opinions.

An interesting direction for future work would be to test our prediction outside mean--field on both real and synthetic networks.  We expect that, unlike the case where the bias in favor of one opinion comes from a broadcast media source~\cite{Colaiori2015b}, the results in the presence of  non--trivial topologies could depend on the location of the zealots, and deviate from mean--field.

\bibliography{BiblioCommittedFinal}

\begin{thebibliography}{21}
\expandafter\ifx\csname natexlab\endcsname\relax\def\natexlab#1{#1}\fi
\expandafter\ifx\csname bibnamefont\endcsname\relax
  \def\bibnamefont#1{#1}\fi
\expandafter\ifx\csname bibfnamefont\endcsname\relax
  \def\bibfnamefont#1{#1}\fi
\expandafter\ifx\csname citenamefont\endcsname\relax
  \def\citenamefont#1{#1}\fi
\expandafter\ifx\csname url\endcsname\relax
  \def\url#1{\texttt{#1}}\fi
\expandafter\ifx\csname urlprefix\endcsname\relax\def\urlprefix{URL }\fi
\providecommand{\bibinfo}[2]{#2}
\providecommand{\eprint}[2][]{\url{#2}}

\bibitem[{\citenamefont{Castellano et~al.}(2009)\citenamefont{Castellano,
  Fortunato, and Loreto}}]{Castellano2009}
\bibinfo{author}{\bibfnamefont{C.}~\bibnamefont{Castellano}},
  \bibinfo{author}{\bibfnamefont{S.}~\bibnamefont{Fortunato}},
  \bibnamefont{and} \bibinfo{author}{\bibfnamefont{V.}~\bibnamefont{Loreto}},
  \bibinfo{journal}{Rev. Mod. Phys.} \textbf{\bibinfo{volume}{81}},
  \bibinfo{pages}{591} (\bibinfo{year}{2009}).

\bibitem[{\citenamefont{Sen and Chakrabarti}(2013)}]{Sen2013}
\bibinfo{author}{\bibfnamefont{P.}~\bibnamefont{Sen}} \bibnamefont{and}
  \bibinfo{author}{\bibfnamefont{B.~K.} \bibnamefont{Chakrabarti}},
  \emph{\bibinfo{title}{Sociophysics: an introduction}}
  (\bibinfo{publisher}{Oxford University Press}, \bibinfo{year}{2013}).

\bibitem[{\citenamefont{Mobilia}(2003)}]{Mobilia2003}
\bibinfo{author}{\bibfnamefont{M.}~\bibnamefont{Mobilia}},
  \bibinfo{journal}{Physical Review Letters} \textbf{\bibinfo{volume}{91}},
  \bibinfo{pages}{028701} (\bibinfo{year}{2003}).

\bibitem[{\citenamefont{Mobilia and G\'{e}orgiev}(2005)}]{Mobilia2005}
\bibinfo{author}{\bibfnamefont{M.}~\bibnamefont{Mobilia}} \bibnamefont{and}
  \bibinfo{author}{\bibfnamefont{I.~T.} \bibnamefont{G\'{e}orgiev}},
  \bibinfo{journal}{Physical Review E - Statistical, Nonlinear, and Soft Matter
  Physics} \textbf{\bibinfo{volume}{71}}, \bibinfo{pages}{1}
  (\bibinfo{year}{2005}), \eprint{0412306}.

\bibitem[{\citenamefont{Mobilia et~al.}(2007)\citenamefont{Mobilia, Petersen,
  and Redner}}]{Mobilia2007}
\bibinfo{author}{\bibfnamefont{M.}~\bibnamefont{Mobilia}},
  \bibinfo{author}{\bibfnamefont{A.}~\bibnamefont{Petersen}}, \bibnamefont{and}
  \bibinfo{author}{\bibfnamefont{S.}~\bibnamefont{Redner}},
  \bibinfo{journal}{Journal of Statistical Mechanics: Theory and Experiment}
  \textbf{\bibinfo{volume}{2007}}, \bibinfo{pages}{P08029}
  (\bibinfo{year}{2007}).

\bibitem[{\citenamefont{Galam and Jacobs}(2007)}]{Galam2007}
\bibinfo{author}{\bibfnamefont{S.}~\bibnamefont{Galam}} \bibnamefont{and}
  \bibinfo{author}{\bibfnamefont{F.}~\bibnamefont{Jacobs}},
  \bibinfo{journal}{Physica A: Statistical Mechanics and its Applications}
  \textbf{\bibinfo{volume}{381}}, \bibinfo{pages}{366 } (\bibinfo{year}{2007}).

\bibitem[{\citenamefont{Xie et~al.}(2011)\citenamefont{Xie, Sreenivasan,
  Korniss, Zhang, Lim, and Szymanski}}]{Xie2011}
\bibinfo{author}{\bibfnamefont{J.}~\bibnamefont{Xie}},
  \bibinfo{author}{\bibfnamefont{S.}~\bibnamefont{Sreenivasan}},
  \bibinfo{author}{\bibfnamefont{G.}~\bibnamefont{Korniss}},
  \bibinfo{author}{\bibfnamefont{W.}~\bibnamefont{Zhang}},
  \bibinfo{author}{\bibfnamefont{C.}~\bibnamefont{Lim}}, \bibnamefont{and}
  \bibinfo{author}{\bibfnamefont{B.~K.} \bibnamefont{Szymanski}},
  \bibinfo{journal}{Physical Review E} \textbf{\bibinfo{volume}{84}},
  \bibinfo{pages}{011130} (\bibinfo{year}{2011}).

\bibitem[{\citenamefont{Marvel et~al.}(2012)\citenamefont{Marvel, Hong, Papush,
  and Strogatz}}]{Marvel2012}
\bibinfo{author}{\bibfnamefont{S.~A.} \bibnamefont{Marvel}},
  \bibinfo{author}{\bibfnamefont{H.}~\bibnamefont{Hong}},
  \bibinfo{author}{\bibfnamefont{A.}~\bibnamefont{Papush}}, \bibnamefont{and}
  \bibinfo{author}{\bibfnamefont{S.~H.} \bibnamefont{Strogatz}},
  \bibinfo{journal}{Phys. Rev. Lett.} \textbf{\bibinfo{volume}{109}},
  \bibinfo{pages}{118702} (\bibinfo{year}{2012}).

\bibitem[{\citenamefont{Yildiz et~al.}(2013)\citenamefont{Yildiz, Ozdaglar,
  Acemoglu, Saberi, and Scaglione}}]{Yildiz2013}
\bibinfo{author}{\bibfnamefont{E.}~\bibnamefont{Yildiz}},
  \bibinfo{author}{\bibfnamefont{A.}~\bibnamefont{Ozdaglar}},
  \bibinfo{author}{\bibfnamefont{D.}~\bibnamefont{Acemoglu}},
  \bibinfo{author}{\bibfnamefont{A.~n.} \bibnamefont{Saberi}},
  \bibnamefont{and}
  \bibinfo{author}{\bibfnamefont{A.}~\bibnamefont{Scaglione}},
  \bibinfo{journal}{ACM Transactions on Economics and Computation}
  \textbf{\bibinfo{volume}{1}}, \bibinfo{pages}{19} (\bibinfo{year}{2013}).

\bibitem[{\citenamefont{Mobilia}(2013)}]{Mobilia2013}
\bibinfo{author}{\bibfnamefont{M.}~\bibnamefont{Mobilia}},
  \bibinfo{journal}{Journal of Statistical Physics}
  \textbf{\bibinfo{volume}{151}}, \bibinfo{pages}{69} (\bibinfo{year}{2013}).

\bibitem[{\citenamefont{Xie et~al.}(2012)\citenamefont{Xie, Emenheiser, Kirby,
  Sreenivasan, Szymanski, and Korniss}}]{Xie2013}
\bibinfo{author}{\bibfnamefont{J.}~\bibnamefont{Xie}},
  \bibinfo{author}{\bibfnamefont{J.}~\bibnamefont{Emenheiser}},
  \bibinfo{author}{\bibfnamefont{M.}~\bibnamefont{Kirby}},
  \bibinfo{author}{\bibfnamefont{S.}~\bibnamefont{Sreenivasan}},
  \bibinfo{author}{\bibfnamefont{B.~K.} \bibnamefont{Szymanski}},
  \bibnamefont{and} \bibinfo{author}{\bibfnamefont{G.}~\bibnamefont{Korniss}},
  \bibinfo{journal}{PLoS ONE} \textbf{\bibinfo{volume}{7}},
  \bibinfo{pages}{e33215} (\bibinfo{year}{2012}).

\bibitem[{\citenamefont{Waagen et~al.}(2015)\citenamefont{Waagen, Verma, Chan,
  Swami, and D'Souza}}]{Waagen2015}
\bibinfo{author}{\bibfnamefont{A.}~\bibnamefont{Waagen}},
  \bibinfo{author}{\bibfnamefont{G.}~\bibnamefont{Verma}},
  \bibinfo{author}{\bibfnamefont{K.}~\bibnamefont{Chan}},
  \bibinfo{author}{\bibfnamefont{A.}~\bibnamefont{Swami}}, \bibnamefont{and}
  \bibinfo{author}{\bibfnamefont{R.}~\bibnamefont{D'Souza}},
  \bibinfo{journal}{Physical Review E} \textbf{\bibinfo{volume}{91}},
  \bibinfo{pages}{1} (\bibinfo{year}{2015}).

\bibitem[{\citenamefont{Svenkeson and Swami}(2015)}]{Svenkeson2015}
\bibinfo{author}{\bibfnamefont{A.}~\bibnamefont{Svenkeson}} \bibnamefont{and}
  \bibinfo{author}{\bibfnamefont{A.}~\bibnamefont{Swami}},
  \bibinfo{journal}{Scientific reports} \textbf{\bibinfo{volume}{5}}
  (\bibinfo{year}{2015}).

\bibitem[{\citenamefont{Mobilia}(2015)}]{Mobilia2015}
\bibinfo{author}{\bibfnamefont{M.}~\bibnamefont{Mobilia}},
  \bibinfo{journal}{Phys. Rev. E} \textbf{\bibinfo{volume}{92}},
  \bibinfo{pages}{012803} (\bibinfo{year}{2015}).

\bibitem[{\citenamefont{Colaiori and Castellano}(2015)}]{Colaiori2015b}
\bibinfo{author}{\bibfnamefont{F.}~\bibnamefont{Colaiori}} \bibnamefont{and}
  \bibinfo{author}{\bibfnamefont{C.}~\bibnamefont{Castellano}},
  \bibinfo{journal}{Phys. Rev. E} \textbf{\bibinfo{volume}{92}},
  \bibinfo{pages}{042815} (\bibinfo{year}{2015}).

\bibitem[{\citenamefont{Strogatz}(2001)}]{Strogatz2001}
\bibinfo{author}{\bibfnamefont{S.~H.} \bibnamefont{Strogatz}},
  \emph{\bibinfo{title}{Nonlinear dynamics and chaos: with applications to
  physics, biology, chemistry, and engineering}} (\bibinfo{publisher}{Westview
  Press}, \bibinfo{year}{2001}).

\bibitem[{\citenamefont{Dodds and Watts}(2004)}]{Dodds2004}
\bibinfo{author}{\bibfnamefont{P.~S.} \bibnamefont{Dodds}} \bibnamefont{and}
  \bibinfo{author}{\bibfnamefont{D.~J.} \bibnamefont{Watts}},
  \bibinfo{journal}{Physical Review Letters} \textbf{\bibinfo{volume}{92}},
  \bibinfo{pages}{218701} (\bibinfo{year}{2004}).

\bibitem[{not({\natexlab{a}})}]{nota0}
\bibinfo{note}{For a given ordinate $y=\overline{y}$,
  $x=(1-\overline{y})(\overline{y}+r)/(r\alpha-\varphi \overline{y})$, so that
  $\partial_r x=-(\varphi+
  \alpha)\overline{y}(1-\overline{y}))/(r\alpha-\varphi \overline{y})^2$, which
  is positive for $\overline{y}$ in the physical region, being
  $-(\varphi+\alpha)=(\beta_1-\varphi_1)/\varphi_2>0$.}

\bibitem[{not({\natexlab{b}})}]{nota3}
\bibinfo{note}{Eliminating the possibility for an individual to be undecided
  leads to a quite trivial behavior: either the system converge to total
  consensus on opinion $A$ (when asymmetric interactions favors $A$), or it
  exhibit a continuous transition between consensus and pluralism (when
  asymmetric interactions favors $B$).}

\bibitem[{\citenamefont{Colaiori et~al.}(2015)\citenamefont{Colaiori,
  Castellano, Cuskley, Loreto, Pugliese, and Tria}}]{Colaiori2015}
\bibinfo{author}{\bibfnamefont{F.}~\bibnamefont{Colaiori}},
  \bibinfo{author}{\bibfnamefont{C.}~\bibnamefont{Castellano}},
  \bibinfo{author}{\bibfnamefont{C.~F.} \bibnamefont{Cuskley}},
  \bibinfo{author}{\bibfnamefont{V.}~\bibnamefont{Loreto}},
  \bibinfo{author}{\bibfnamefont{M.}~\bibnamefont{Pugliese}}, \bibnamefont{and}
  \bibinfo{author}{\bibfnamefont{F.}~\bibnamefont{Tria}},
  \bibinfo{journal}{Physical Review E} \textbf{\bibinfo{volume}{91}},
  \bibinfo{pages}{012808} (\bibinfo{year}{2015}).

\bibitem[{not({\natexlab{c}})}]{nota4}
\bibinfo{note}{Examples are systems in the FCM class that have $\gamma_1 <
  \varphi_1$ and $\varphi_2>\gamma_2$ (the first condition means that more $B$
  than $A$ states are lost in $A-B$ interactions; the second that individuals
  in $A$ state are more successful than those in $B$ state in convincing
  undecided individuals). In this case, both the presence of zealots and the
  rules of peer interactions favor A, yet for small densities of zealots and
  suitable initial conditions the system reaches a stationary state with finite
  density of dissenters.}

\end{thebibliography}

\end{document}